\documentclass{mem}
\usepackage{natbib}\usepackage{txfonts}\usepackage{balance}
\usepackage{flushend}
\usepackage{graphicx}
\usepackage[breaklinks,pdftex]{hyperref}
\idline{75}{282}
\begin{document}
\def\teff{$T\rm_{eff }$}
\def\kms{$\mathrm {km s}^{-1}$}

\title{The chemistry of star and planet formation with SKA}


\author{
C. Codella\inst{1} \and L. Testi\inst{2} 
\and  G. Umana\inst{3}  \and S. Molinari\inst{4} \and E. Bianchi\inst{5,1}
\institute{INAF, Osservatorio Astrofisico di Arcetri, Largo E. Fermi 5, I-50125, Firenze, Italy
\and
Dipartimento di Fisica e Astronomia Augusto Righi, Universit\`a di Bologna, Viale Berti Pichat 6/2, Bologna, Italy
\and
INAF – Osservatorio Astrofisico di Catania, Via S. Sofia 78, I-95123 Catania, Italy
\and
INAF-Istituto di Astrofisica e Planetologia Spaziali, Via del Fosso del Cavaliere 100, 00133 Rome, Italy
\and 
Excellence Cluster ORIGINS, Boltzmannstraße 2, 85748, Garching bei Mu\"unchen, Germany}\\
\email{claudio.codella@inaf.it}
}

\authorrunning{Codella et al.}

\titlerunning{Astrochemistry with the SKA}

\date{Received: XX-XX-XXXX (Day-Month-Year); Accepted: XX-XX-XXXX (Day-Month-Year)}

\abstract{
In this contribution, we aim to summarise the efforts of the Italian SKA scientific community in conducting surveys of star-forming regions within our Galaxy, in the development of astrochemical research on protostellar envelopes and disks, and in studying the planet formation process itself. 
The objective is dual: Firstly, to investigate the accumulation and development of dust throughout the formation of planets, and secondly, to chemically examine protoplanetary disks and protostellar envelopes by studying heavy molecules, such as chains and rings containing over seven carbon atoms, which exhibit significantly reduced strength at millimeter wavelengths.
}
\maketitle{}

\section{Introduction}\label{sec:intro}

The ingredients for the recipe to make a habitable planet like our own Earth are: a small rocky planet at the right distance from the host star for water to be in the liquid state, a magnetosphere, and an atmosphere organic-rich in volatiles, capable of developing organic molecules chemistry. Planets form in protoplanetary disks in the early phases of stellar evolution from the material processed and left over during the star formation process. 
In the last few decades, the field of exoplanets has shown a large degree of diversity in the planetary systems, and one of the main modern challenges of the planet formation field is to identify the common traits, physical and chemical processes shared by all systems. 
Studying the chemical and physical evolution of the refractory (dust) and volatile (gas) component of the interstellar medium (ISM) during the process of star and planet formation is thus key to understand the genesis of our own Solar System, the development of life on Earth, and eventually how common or peculiar our system is in the global context.
Key questions still to be addressed are: how chemically organic complex are the volatiles delivered to the pristine planetary atmospheres? What molecules are passed from the large-scale envelope to the disk where planets, comets, and asteroids form? What is the role of the star and environment in influencing this chemical evolution? These are key questions in the context of the SKA Cradle of Life\footnote{\url{https://www.skao.int/index.php/en/science-users/science-working-groups/105/cradle-life}} working group, which is dedicated to developing scientific cases for the SKA in the fields of star and planet formation and evolution, including astrochemistry. 
The main lines of research regard the detection and characterisation of large molecules in the planet-forming regions, the study of grain growth to constrain the planet formation process, the detection of magnetic fields around exoplanets via auroral radio emission, and stellar activity, the study of Solar System objects, and the search for extraTerrestrial Intelligence (SETI).
 
Preparatory work inside the Cradle of Life working group includes pilot projects with SKA pathfinders and precursors \citep[e.g.][]{Coutens2019,Bianchi2023a} and a constant upgrade of the SKA1 use cases\footnote{\url{https://www.skao.int/sites/default/files/documents/d35-SKA-TEL-SKO-0000015-04_Science_UseCases-signed.pdf}}. 


\section{SKA technical specifications}
\label{sec:grazia}

The Square Kilometer array (SKA) project is an international effort to build the largest radio telescope in the world, with a collecting area of more than one square kilometer.
The SKA Observatory, a new intergovernmental organization dedicated to radio astronomy, headquartered in the United Kingdom, at the University of Manchester's Jodrell Bank Observatory, has responsibility for coordinating the overall project and managing two remote telescopes. 
The SKA has a low-frequency element, the SKA-LOW telescope, located at the Murchison Radio Observatory in Western Australia, and a high-frequency element, the SKA-MID telescope, located in the Karoo region of South Africa.  Both sites were chosen for their optimal atmospheric conditions above the desert sites andfor their
radio quietness, among many other scientific and technical reasons. 

\begin{figure*}
\begin{center}
    \includegraphics[scale=0.30,angle=0]{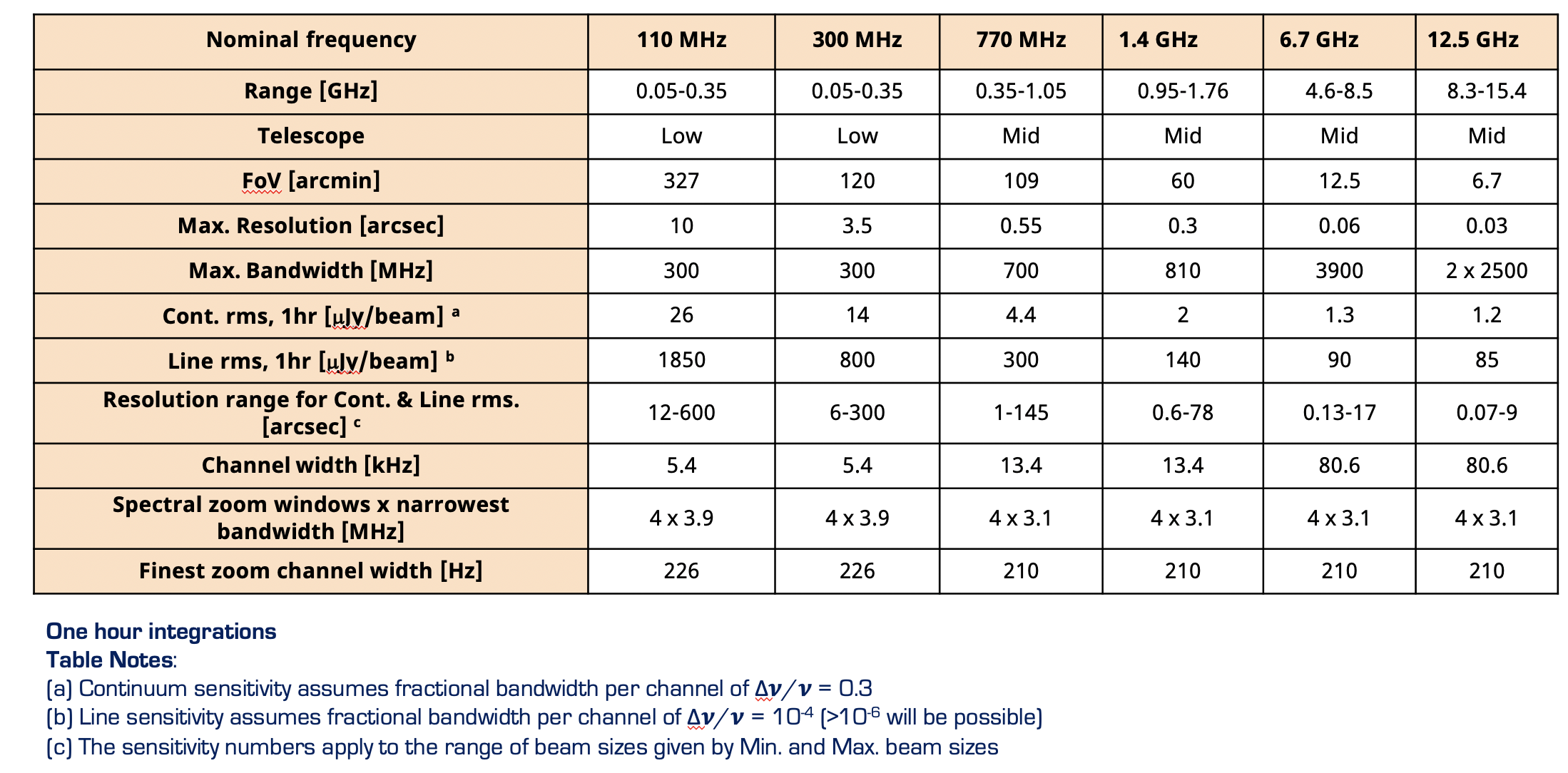}
\caption{\footnotesize Summary of the anticipated imaging performance of SKA.
Adapted from \cite{Braun2019}.}
\label{fig:Table1}
\end{center}
\end{figure*}

SKA-LOW, currently under construction, will consist of 512 aperture array stations, each composed of 256 dual-polarised log-periodic antennas randomly distributed in a 38 m diameter circular area (131,072 antennas in total). The inner aperture array stations are arranged in a compact core with a diameter of 1 km with three spiral arms with a maximum baseline of 65 km. The antenna array will operate from 50 MHz to 350 MHz. 
SKA-MID, currently under construction, will consist of an array of 197 reflector antennas. This will be a mixed array as the 64, 13.5m diameter, dishes from MeerKAT telescope will be fully integrated to the final array of 133,  15m diameter,  SKA dishes. The final configuration will have a 150-km baseline. The outer part of the array is a 3-arm spiral configuration, providing excellent instantaneous coverage of the u-v  plane.  The dense core provides excellent sensitivity to extended regions of low brightness and a sensitive (single-dish-like) sub-array for pulsar and transient astronomy. The antenna array will operate from 350 MHz to 15.3 GHz.

\begin{figure}
\begin{center}
    \includegraphics[scale=0.23,angle=0]{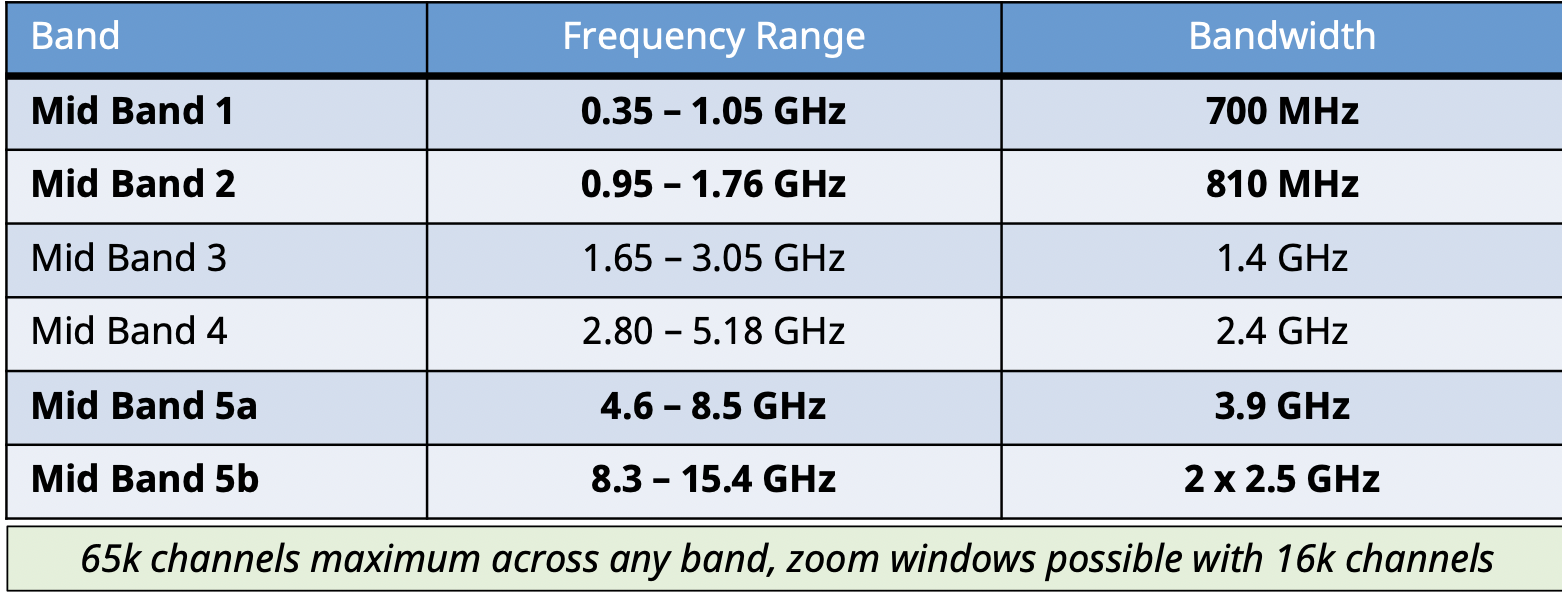}
\caption{\footnotesize SKA-MID frequency coverage. Bands in bold will be deployed as part of the funded Design Baselines and were chosen after a science prioritisation (see text).}
\label{fig:Table2}
\end{center}
\end{figure}

\begin{figure*}
\begin{center}
    \includegraphics[scale=0.30,angle=0]{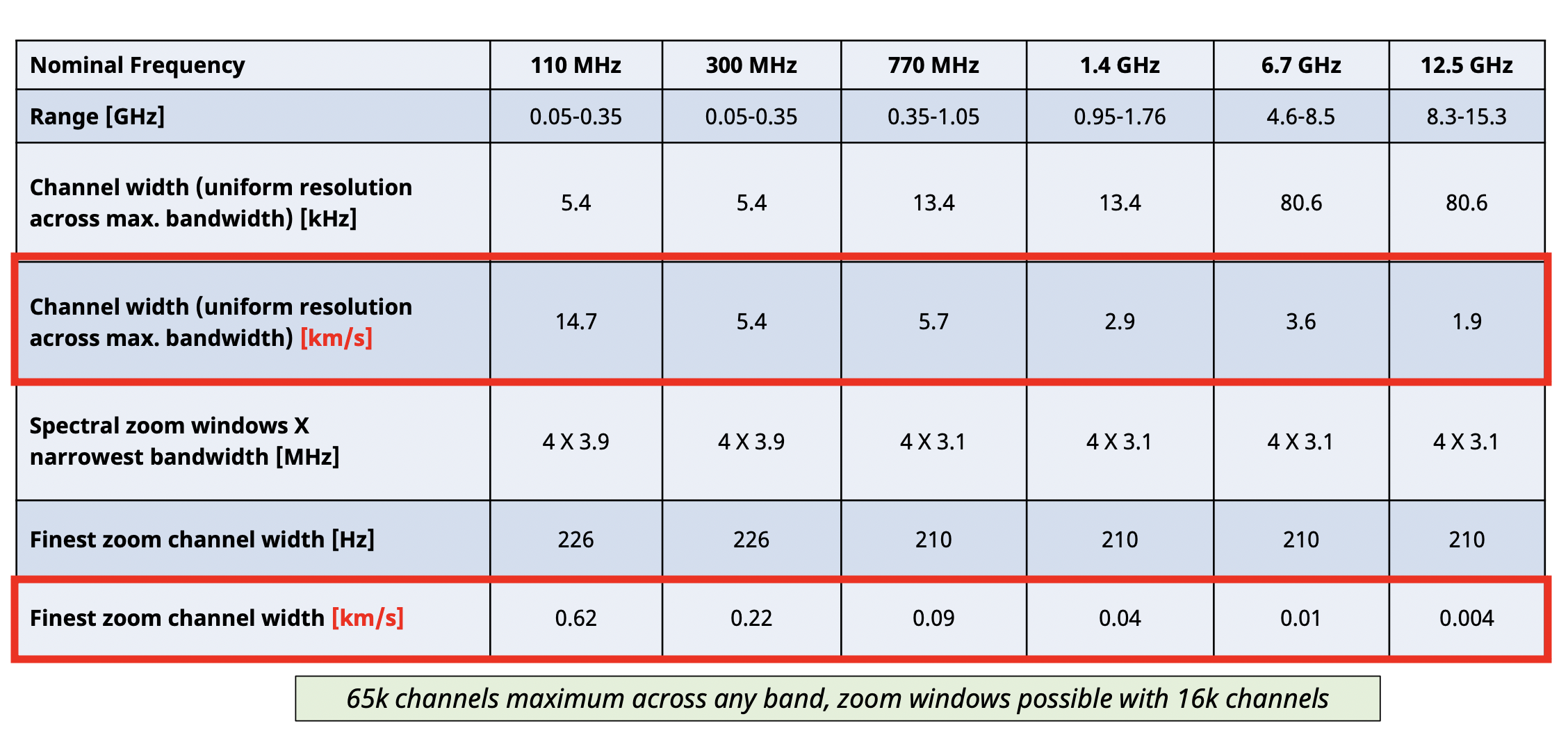}
\caption{\footnotesize SKA spectral resolutions (red boxes point out the estimates in km s$^{-1}$) at different frequencies.}
\label{fig:Table3}
\end{center}
\end{figure*}

Figure \ref{fig:Table1}, adapted from \cite{Braun2019}, provides a summary of the anticipated imaging performance of SKA. The frequency coverage and bandwidth of SKA-MID are listed in Fig. \ref{fig:Table2}.  Bands in bold will be deployed as part of the funded Design Baselines and were chosen after a science prioritization\footnote{\url{https://skao.canto.global/s/M8159?viewIndex=0&column=document&id=bg07p5lsdt2ep2iv9660tpd701}}.
Bands 3 and 4, that are part of the Design Baseline, are not currently funded. While the frequency coverage of the Design Baseline for SKA-MID stops at 15.3 GHz, the dishes have an aperture efficiency specification for 20 GHz, with possible good performance up to at least 25 GHz for possible future expansion in frequency space (i.e. Band-6).
Spectral line imaging provides between 52,500 and 65,536 linearly spaced channels across the frequency band. Zoom windows allow for higher spectral resolution images to be obtained, down to several hundreds of Hz. These windows contain between 14,000 and 16,384 linearly spaced channels that can be configured across the available bandwidth. Each zoom window of 3.1 MHz contains 16k channels.  The maximum zoom of the 5a/b Band results in very fine velocity channels (0.004 km/sec at 12.5 GHz). Details on channel widths at other frequencies are summarised in Fig. \ref{fig:Table3}.

\section{Complex carbon chains and rings at the onset of planet formation}
\label{sec:eleonora}

\begin{figure}[t!]
\resizebox{\hsize}{!}{\includegraphics[clip=true]{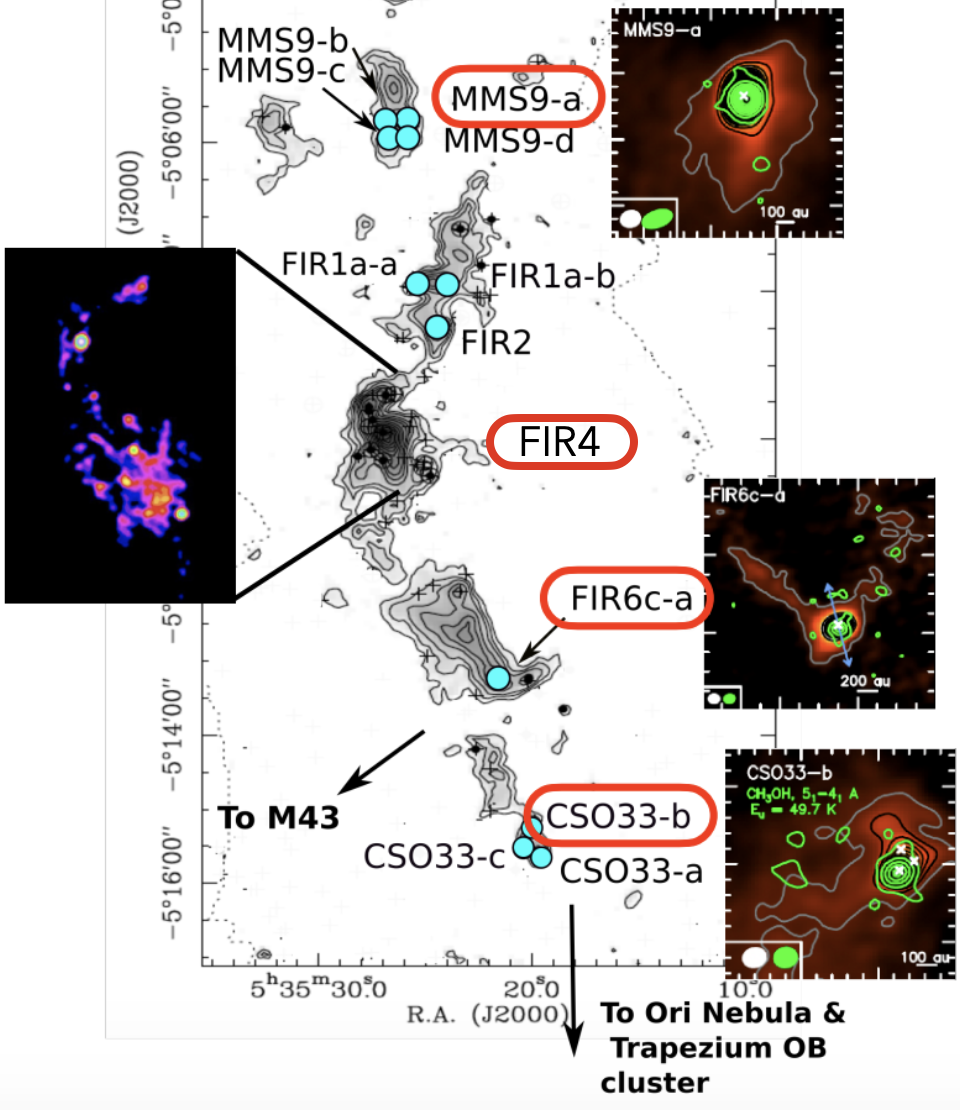}}
\caption{\footnotesize
The OMC2/3 star forming region observed by the ALMA ORANGES survey. Adapted from \citet{Bouvier2022}.
}
\label{fig:orion}
\end{figure}

Life on Earth, as we understand it, is carbon-based, with carbon acting as the primary structural foundation, or "backbone," of prebiotic molecules \citep{Ceccarelli2023, Oberg2021, Mumma2011}. Despite their pivotal role in synthesizing prebiotic compounds, complex carbon chains and rings (molecules with more than 5 carbon atoms) remain relatively unexplored in astrochemical studies due to their faint spectral lines at mm-wavelengths.
Complex carbon species are abundant and widespread in the early stages of Solar-type stars, and they are mostly detected from single-dish observations \citep[][]{Sakai&Yamamoto}. 
Recent observations have highlighted the importance of hydrocarbons and complex carbon species, not only in astrobiology but also in the chemistry of young Solar analogues. Specifically, the detection of complex species such as C$_4$H, C$_6$H, HC$_7$N, HC$_9$N, c-C$_9$H$_8$, o-C$_6$H$_4$, and c-C$_6$H$_5$CN in TMC-1 and other prestellar sources has demonstrated the prevalence of a rich aromatic carbon chemistry during the earliest stages of star formation \citep{McGuire2020,Burkhardt2021, Cernicharo2021, Bianchi2023a}. Complex carbon species have been detected in protostellar sources \citep[e.g.][]{Sakai2008} and protostellar streamers \citep{Taniguchi2024-streamer}. Recent JWST observations have confirmed an active warm hydrocarbon chemistry in the inner disk around a low-mass star \citep{Tabone2023}. Although complex carbon species are detected in various environments, their chemical origin remains poorly understood as well as their spatial distribution at small spatial scales. The SKA will provide for the first time observations of complex carbon species, which have their emission peak at radio frequencies, in the planet-formation region. To this end a dedicated science case has been developed in the context of Cradle of Life\footnote{\url{https://www.skao.int/sites/default/files/documents/d35-SKA-TEL-SKO-0000015-04_Science_UseCases-signed.pdf}}. The focus of this project will be to observe with SKA the Orion Molecular Cloud 2 (OMC-2) star-forming region. This active star-forming filament harbors a diverse population of both low- and high-mass stars and disks \citep[e.g.][]{Tobin2019}, making it suitable for observing multiple objects within the large field of view ($>$ 6$\arcmin$) of the SKA. Among these, the FIR 4 region stands out due to its exposure to a flux of high-energy cosmic-ray-like particles \citep{Ceccarelli2014, Fontani2017}. This heightened exposure closely resembles the conditions encountered by the early Solar System, which formed within a dense cluster of stars \citep{Lichtenberg2019}. As a result, OMC-2 FIR4 is regarded as one of the closest analogs to the environment in which our Sun may have formed, making it an ideal location for studying chemistry reminiscent of our early Solar System.
Thanks to the unprecendent angular resolution and sensitivity provided by the SKA we will detect in Band 5 (5--15 GHz) emission lines from complex carbon chains and rings at spatial scales of 200 au around disks in OMC-2.
This will allow for the first time to have a complete census of the chemical complexity available for planet formation, complementing large astrochemical surveys performed at (sub-)mm wavelenghts. Dedicated large programs performed with single-dish telescopes, such as ASAI \citep{Lefloch2018} with the IRAM-30m, and interferometers, such as SOLIS \citep{Ceccarelli2017} performed with IRAM-NOEMA and FAUST \citep{Codella2021} with ALMA, have revealed a rich chemistry including complex organic molecules and potentially prebiotic species \citep{Ceccarelli2023}. Interestingly, the ALMA ORANGES survey has also revealed a different incidence of sources rich in methanol, the precursor of several complex organic molecules, in the OMC2/3 region with respect to Perseus (see Figure \ref{fig:orion}, \citealt{Bouvier2022}). In particular, the analysis of 19 solar-mass protostars showed that only 26\% had warm methanol emission compared to 60\% in Perseus (\cite{Yang2021}.
This difference may be related to differences in the cloud conditions, and in particular to the proximity with the Trapezium cluster providing an intense UV illumination in the southern part of the filament. Observations performed with SKA will include protostellar objects located at different distances from the Trapezium cluster, allowing us to investigate the feedback mechanisms including protostellar shocks
and outflows, external UV illumination from nearby stars and local cosmic rays, which can extract carbon atoms from CO, making them available for the formation of carbon species \citep{Bianchi2023a}. In addition to protostellar feedback, a short duration of collapse timescales during the prestellar core stage may also play an important role in favoring the formation of complex carbon chains and rings. This is because it will enrich icy dust grain mantles with CH$_4$ instead of CO \cite{aikawa_chemical_2020}.
The proposed SKA observations will identify the main mechanism of complex carbon species formation, shedding light on the origins of the chemical diversity in Solar System precursors. This includes protostellar systems enriched with complex organic molecules, while others show no emission from these species and instead exhibit warm carbon chain chemistry \citep{Ceccarelli2023, Sakai&Yamamoto} and hydrocarbons \cite{Tabone2023}. Understanding the origin of this chemical diversity observed at early evolutionary stages will bring us a step closer to comprehending the variety of exoplanets and the origin of Life in our Solar System.

\section{Dust evolution during and after planet formation}
\label{sec:leonardo}

The evolution of dust plays a pivotal role in the planet formation process as this is the material that will assemble into e.g. planetesimals and then minor bodies, and the bulk of the terrestrial planets. 
Grain settling and growth in protoplanetary disks is thought to be the initial step of the formation of the rocky cores of planets \citep[see e.g.][]{2000prpl.conf..533B}. The processes of collision and sticking of grains have been studied theoretically and in laboratory experiments \citep[eg][]{2008ARA&A..46...21B}, and are used in global disk evolution models to describe the physics of dust evolution \citep[eg.][]{2012A&A...539A.148B}. The review of \citet{2014prpl.conf..339T} provides a summary of the theoretical, experimental and observational understanding of the grain growth process in disks. 

The main observational signature for the growth of grains in the disks midplane is the spectral index of the dust opacity coefficient, which can be measured once the thermal structure of the disk is modeled and contamination from other sources of continuum emission (e.g. from electrons in winds, or the non-thermal emission from the central young star) is taken into account \citep{2001ApJ...554.1087T,2003A&A...403..323T}. The combination of sensitive mm- and cm-observations of disks in nearby star forming regions allows us to disentangle emission mechanisms and provide evidence for significant grain growth \citep{2021MNRAS.506.2804T}. Nevertheless, the discovery of substructures in disks, thanks to ALMA sensitivity and angular resolution \citep[e.g. HL Tau][]{2015ApJ...808L...3A} has allowed us to realize that the effect of dust segregation in narrow rings and, consequently, the local increase in the emission optical depth is a significant limitation for our efforts to probe the level of grain growth. Detailed multi-wavelength analysis of well resolved disks show that, while evidence for grain processing is confirmed, determining the exact level of growth and dust/gas ratio enhancement in the ring is difficult \citep[e.g.][]{Carrasco2016,Carrasco2019,2019MNRAS.486.4638U,2022A&A...665A..25C,2022A&A...664A.137G}. 

One of the latest development in the field is related to the recognition of the importance of disk planet interaction in shaping the evolution of solids in disks. Theoretical considerations and observational constraints seem to converge on the idea that giant planets form relatively fast in disks, which is confirmed by the apparent lack of enough dust mass to form cores of planets in disks older than 1-2 Myr \citep[eg][]{2018A&A...618L...3M,2016A&A...593A.111T,2022A&A...663A..98T}. The apparent lack of dust evolution in disks at these ages may be related to the fact that disk-planet interaction promote dust production via planetesimal collisions \citep[see e.g.][]{2019ApJ...877...50T,2022ApJ...927L..22B}. This view appears to be confirmed by the discovery of massive planets in disks via the disturbances they introduce in the disk kinematics \cite[][]{2022ApJ...928....2I,2022A&A...665A..25C,2023A&A...674A.113I}.

If planet formation occurs at early stages, the first evidences for dust evolution need to appear early on, during the disk formation phase. Indeed, a recent development in the field has been the analysis of dust properties in protostellar envelopes and disks \citep{2019A&A...632A...5G,2023A&A...676A...4C}, supporting an early dust processing. In the coming years, ALMA and NOEMA will continue to characterize the dust properties in protostellar envelopes and disks at millimetre wavelengths, also employing the new receivers of the Wideband Sensitivity Upgrade \citep[e.g.][]{2020A&A...634A..46Y} and polarization capabilities, which provide complementary constraints on the dust evolution \citep{2019MNRAS.488.4897V}. Nevertheless, optical depth will remain a major obstacle in using millimetre wavelenths observations to characterize the dust evolution in disks and protostars. 

Which will be the role of SKA in this science domain?
SKA will allow us to achieve unprecedented sensitivity at longer wavelengths, overcoming the limitations imposed by optical depth and will allow us to probe directly the planet forming region ($\le 10$ au) in protostars. A detailed account of the SKA contribution in this area is provided in the SKA Science Book \citep{2015aska.confE.117T}. Waiting for SKA, our community has started to characterise the long wavelength emission of disks with the SKA precursors: a major effort has been dedicated to the study of the $\rho$-Ophiuchi core with the VLA \citep{Coutens2019}, this study will need to be followed up in other star forming regions and at other frequencies, overlapping with the ALMA and VLA surveys. These data will characterise the properties of the radio emission from young stars as a function of age and stellar parameters, this is a critical step to understand the effect of stellar activity on the chemical evolution of the solid and volatile material during the planet formation process, for example the photoevaporation of the disk by energetic radiation from the central star, which can be indirectly traced via the radio continuum emission \citep{2012ApJ...751L..42P,2014A&A...570L...9G}. An essential ingredient to understand the circulation of processed material in the planet forming zone is the role of winds and outflows in the radial advection processes. There is tentative evidence for a correlation between dust properties and the strength of winds in young stars and disks \citep{2024ApJ...961...90C}, as well as a correlation between outflows and disk star interaction in the form of accretion (Garufi et al. submitted). 
SKA will allow us to finally provide final answers to many of these pressing questions in planet formation. Some specific contributions to some of these topcis are discussed in this book in Chapter~4 (Perotti et al.). 



\section{SKA beyond 15 GHz: iCOMs in disks as observed in Band 6}
\label{sec:claudio2}

\begin{figure}
\begin{center}
    \includegraphics[scale=0.33,angle=0]{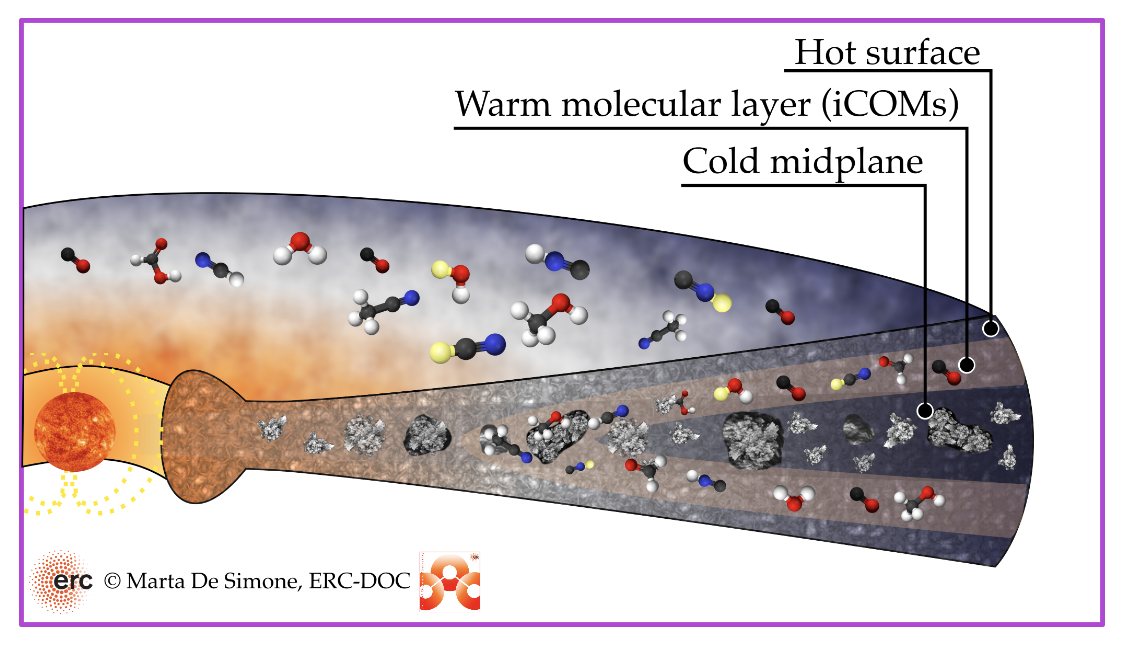}
\caption{\footnotesize
Sketch of a typical protoplanetary disk. The heated outer layer is where UV light ionises atoms and breaks apart molecules. In the warm molecular layer, which captures the UV light, gaseous molecules (e.g. iCOMs) are stable. Within cold midplane, almost all elements solidify onto the dust surfaces. From \citet{Ceccarelli2023}.}
\label{fig:disk}
\end{center}
\end{figure}

Numerous studies have indicated that the chemical makeup at the onset of a Solar-type planetary system's formation serves as a crucial indicator for tracing its developmental history
\citep[e.g.][]{caselli_2012}. Out of the mre than 310 molecules so far identified in the ISM, interstellar complex organic molecules (iCOMs) are particularly noteworthy. They are not only valuable for their ability to trace evolutionary pathways but also because they may be the forerunners of biomolecules \citep[see e.g.][]{Saladino2012}, the building blocks of life on Earth. The emission of iCOMs has been revealed during all 
the initial phases of a Solar-type planetary system's development, which includes the prestellar, protostellar environments, and protoplanetary disc stages \citep[see the recent review by][]{Ceccarelli2023}. Within this framework, the significance of discs stems from their role as the birthplace of planets. 

Thermo-chemical models suggest that protoplanetary disks surrounding young stars similar to the Sun (see Fig. \ref{fig:disk}) are composed of three distinct chemical layers
\citep[see e.g. the review by][]{Oberg2023}:
(1) The hot surface layer, also known as the disk atmosphere, is where intense radiation breaks down molecules through photodissociation; (2) The warm molecular layer is characterized by the presence of molecules in their gaseous state, with active gas-phase chemical reactions; (3)  The freeze-out layer refers to the cold midplane of the outer disk, where molecules adhere to dust particles and form icy mantles. For each molecule, the freeze-out occurs at the disk radius and height where the dust
temperature falls below the freeze-out temperature, which depends on the molecular binding energy. Nonetheless, nonthermal desorption mechanisms, triggered by the penetration of UV and X-rays from stars, can cause molecules to freeze out beyond their expected snowlines, occurring at greater heights within the outer disk \citep{Walsh2014}.
Observations of disks conducted at wavelengths under 1mm 
are affected by optically thick dust, obscuring line emission from the inner disk region. 

\begin{figure}
\includegraphics[scale=0.5,angle=0]{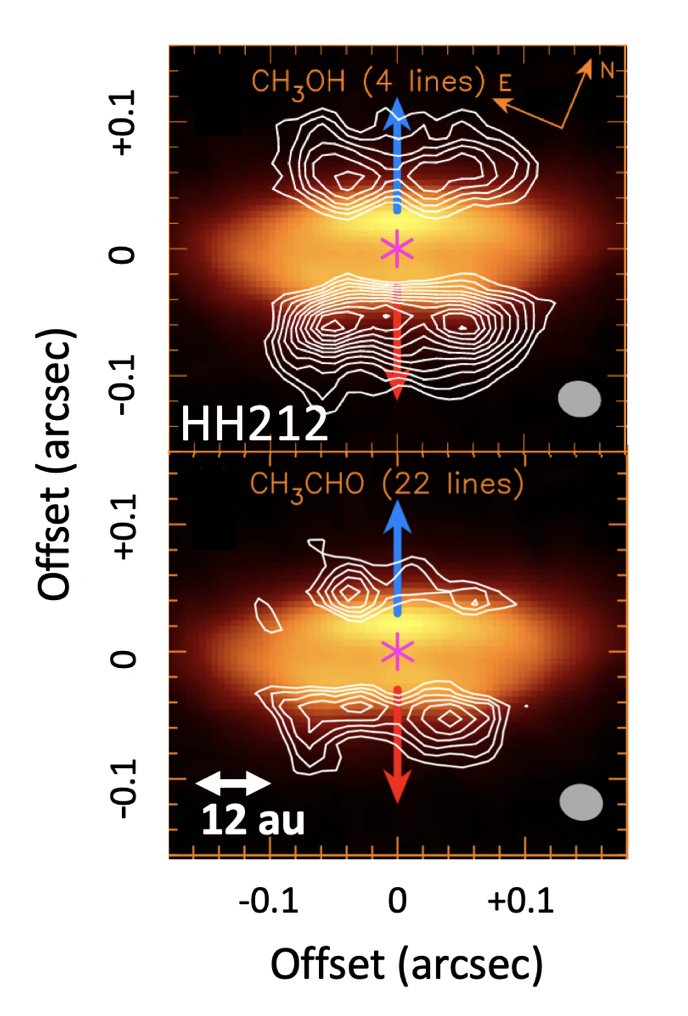}
\caption{\footnotesize
Methanol and acetaldehyde 
line emission (white contours, ALMA-Band 7 observations) overlapped to the dust continuum emission (color image) towards the protostellar disc around the Class 0 HH 212-mm object. The blue and red arrows indicate the axes of the blue- and red-shifted jet components, respectively. Adapted from \citet{Lee2019}.}
\label{fig:hh212}
\end{figure}

The case of the protostellar disc around the Class 0 HH212-mm (Orion) as observed with ALMA at sub-millimeter wavelengths is instructive \citet{Codella2019}. Figure \ref{fig:hh212} reveals the spatial distribution of the chemically-enriched gas around HH 212-mm,
showing for the first time \citep{Lee2019,lee_2022} 
two rotating outer layers of the disk (here in CH$_3$OH and CH$_3$CHO emission, but several complex organics, such as methyl formate, were
imaged). The key point in this context is that no iCOMs 
emission is detected toward the dusty disk within 0$\farcs$1 (40 au) of the center, where the dust emission in ALMA-Band 7 is severely optically thick. The lack of emission toward the disk midplane is clearly seen, but what is not clear is the origin of such a morphology. 
This could be due to the previously mentioned dust optical depth effects or a lower abundance of these molecules in those areas. To accurately determine the intricate molecular makeup of the midplanes of protoplanetary disks (where planet formation occurs), it is crucial to conduct observations at lower frequencies, specifically at centimeter wavelengths.

SKA-Band 6 would offer the best combination to optimise the dynamical range of the expected emission lines due to iCOMs. Frequencies exceeding approximately 15 GHz encompass transitions that are up to 500 times more intense than those found in Band 5, providing an optimal opportunity for the detection and characterization of these molecules close to the equatorial plane of protoplanetary discs using the SKA \citep{Walsh2014}. 
Further details in this context can be found in the SKA MEMO 20-01 (\url{www.skatelescope.org/memos}; SKA1 Beyond 15 GHz:
The Science case for Band 6) document, and, more precisely, in the contribution by J.D. Ilee, C. Codella, C. Walsh and collaborators\footnote{www.skao.int/sites/default/files/documents/d38-ScienceCase$\textunderscore$band6$\textunderscore$Feb2020.pdf}.

\begin{acknowledgements}
CC, and EB acknowledge the PRIN-MUR 2020  BEYOND-2p (Astrochemistry beyond the second period elements, Prot. 2020AFB3FX),
the PRIN MUR 2022 FOSSILS (Chemical origins: linking the fossil composition of the Solar System with the chemistry of protoplanetary disks, Prot. 2022JC2Y93), the project ASI-Astrobiologia 2023 MIGLIORA
(Modeling Chemical Complexity, F83C23000800005), the INAF-GO 2023 fundings
PROTO-SKA (Exploiting ALMA data to study planet forming disks: preparing the advent of SKA, C13C23000770005), and the INAF Mini-Grant 2022 “Chemical Origins”. 
E.B. acknowledges support from the Deutsche Forschungsgemeinschaft (DFG, German Research Foundation) under Germany´s Excellence Strategy – EXC 2094 – 390783311. 
This work was partly supported by the Italian Ministero dell’Istruzione, Uni- versità e Ricerca through the grant Progetti Premiali 2012-iALMA (CUP C52I13000140001), and has received funding from the European Union’s Horizon 2020 research and innovation program under the Marie Skłodowska Curie grant agreement No. 823823 (DUSTBUSTERS), and from the European Research Council (ERC) via the ERC Synergy Grant ECOGAL (grant 855130)
This paper makes use of the following ALMA data: ADS/JAO.ALMA\#2018.1.01205.L (PI: S. Yamamoto). ALMA is a partnership of the ESO (representing its member states), the NSF (USA) and NINS (Japan), together with the NRC (Canada) and the NSC and ASIAA (Taiwan), in cooperation with the Republic of Chile. The Joint ALMA Observatory is operated by the ESO, the AUI/NRAO, and the NAOJ.
\end{acknowledgements}

\bibliography{astrochemistry.bib}{}
\bibliographystyle{aa} 
\end{document}